\documentclass[twocolumn,showpacs,showkeys,preprintnumbers,amsmath,amssymb]{revtex4}
\usepackage{amsfonts}
\usepackage{amssymb}
\usepackage{color}
\usepackage{txfonts}
\usepackage{graphicx}
\usepackage{dcolumn}
\usepackage{bm}
\usepackage{cases}
\usepackage{amssymb}
\usepackage{txfonts}
\usepackage{graphicx}
\usepackage{dcolumn}
\usepackage{bm}
\usepackage{amssymb}
\usepackage{amsmath}
\usepackage{epsfig}
\usepackage{multirow}
\usepackage{threeparttable}
\begin{document}
\preprint{}
\title{Improvement in corrosion resistance and biocompatibility of AZ31 magnesium alloy by NH$_{2}^{+}$ ions}
\author{Xian Wei\textsuperscript{1,2}}
\author{Zhicheng Li\textsuperscript{3}}
\author{Pinduo Liu\textsuperscript{4}}
\author{Shijian Li\textsuperscript{1}}
\author{Xubiao Peng\textsuperscript{1}}
\author{Rongping Deng\textsuperscript{5}}
\author{Qing Zhao\textsuperscript{1}}

\email[The corresponding author Email: ]{qzhaoyuping@bit.edu.cn}

\affiliation{\textsuperscript{1}School of Physics, Beijing Institute of Technology, Beijing 100081, China}
\affiliation{\textsuperscript{2}Department of science, Taiyuan Institute of Technology, Taiyuan 030008, China}
\affiliation{\textsuperscript{3}Department of Physics, Taiyuan Normal University, Taiyuan 030031, China}
\affiliation{\textsuperscript{4}School of Life Science, Beijing Institute of Technology, Beijing 100081, China}
\affiliation{\textsuperscript{5}Science Division, Beloit College, 700 College Street, Beloit, Wisconsin 53511, USA}

\begin{abstract}
Magnesium alloys have been considered to be favorable biodegradable metallic materials used in orthopedic and cardiovascular applications.
We introduce NH$_{2}^{+}$ to the AZ31 Mg alloy surface by ion implantation at the energy of 50 KeV with doses ranging from $1\times10^{16}$ ions/cm$^{2}$ to $1\times10^{17}$ ions/cm$^{2}$ to improve its corrosion resistance and biocompatibility.
Surface morphology, mechanical properties, corrosion behavior and biocompatibility are studied in the experiments.
The analysis confirms that the modified surface with smoothness and hydrophobicity significantly improves the corrosion resistance and biocompatibility while maintaining the mechanical property of the alloy.
\end{abstract}

\keywords{Magnesium alloy; Corrosion resistance; Biocompatibility; Ion implantation}
\maketitle

\section{Introduction}
Magnesium and its alloys as biodegradable metals have received much attention due to their promising applications in orthopedic, cardiovascular and other medical fields \cite{ErbelW, WaksmanH, WitteW, KaeseW, CesnjevarW}.
For example, biodegradable implants can provide mechanical support during the mending of injured tissue and can also be decomposed in the body fluid after cure rather than been removed by a second surgery \cite{JinC, WitteF}.
The magnesium ions released from the implanted material also offer beneficial effects on bone growth \cite{GuZ, ZhaoJ, FeyerabendH}.
Moreover, magnesium and its alloys have remarkable advantages over the Ti-based alloys and stainless steels in density and elastic modulus.
The density of magnesium alloys range from 1.74 to 2.0 g/cm$^{3}$ and the elastic modulus is from 41 to 45 GPa which are closer to those of human bones compared to the Ti-based alloys (4.4-4.5 g/cm$^{3}$ and 110-117 GPa) and stainless steels (7.9-8.1 g/cm$^{3}$ and 189-205 GPa) \cite{StaigerD, ali2019magnesium}.

There are still a number of challenges in the development of magnesium alloys for medical applications.
Since the intrinsic standard corrosion potential of Mg (-2.37 $V_{SCE}$) is smaller than that of other biodegradable metals (Zn: -0.76 $V_{SCE}$, Fe: -0.44 $V_{SCE}$) \cite{Scheideler, Y.Zheng}, pure magnesium and most magnesium alloys have poor performance in corrosion resistance and produce hydrogen gas during the degradation process \cite{SankarC}.
Additionally, unlike applications in pyrotechnics, metallurgical industry, and automotive products, biomedical field requires magnesium and its alloys to possess not only controlled degradation rate but also excellent biocompatibility \cite{LiH, KainerKU}.
Therefore, a lot of methods such as alloying, coating and surface treatment have been used to enhance both the corrosion resistance and the biocompatibility of magnesium and its alloys.

Ion implantation is one of the effective surface treatment methods to improve the anticorrosion and biocompatibility of magnesium alloys through the surface modification.
Ion implantation can form a functional surface by inserting a beam of varied ionized particles;
it also has the following outstanding advantages: 1) It does not change the size and shape of the samples, so it can be used in the last step in the product manufacturing process, especially for those small, complex and irregular parts (such as screws, nuts, etc.);
2) The implanted ions have high purity and controllable concentration;
3) The treatment process is clean and environmental friendly.
Many types of ions, such as gaseous ions O$_{2}$ \cite{WuGS, WanG.J.}, N$_{2}$ \cite{GuosongWu, ShinjiF, WuDing}, metallic ions Gd \cite{TaoXW}, Ce \cite{WangXM}, and dual ions Zr-N\cite{ChengQiao}, Cr-O \cite{XuRZWu}, have been utilized in the ion implantation technology to effectively improve the corrosion resistance of magnesium and its alloys.
Additionally, a class of organic functional ion implantation has been applied in other fields such as chemical modified electrode, graphene, and bioceramic \cite{ZhangMXGu, GaoDMSun, LiEffects, GuoNcontain}.
Zhao et al. \cite{Zhaoqing} reported that Al$_{2}$O$_{3}$ ceramics bombarded with NH$_{2}^+$ ions demonstrated excellent performance in biocompatibility with living bones in the animal tests.
As Li et al. \cite{LiDJNiulf} reported that the 3T3 mouse fibroblasts and human endothelial cells achieved better attachment and proliferation cultured on the surface of the polypropylene implanted by COOH$^+$ ions.
However, to the best of our knowledge, there is limited information about the organic functional ions in Mg alloy surface modification.
In the consideration of the potential of ion implantation for anticorrosion and the less toxicity of organic molecules comparing to the metal ions, this paper reports a study of improving the corrosion resistance and the biocompatibility of AZ31 Mg alloy by NH$_{2}^{+}$ implantation.
The mechanical performance of the alloy is measured by Nano indenter.
The corrosion property is analyzed by both electrochemical and immersion tests, and the biocompatibility of the modified alloy is investigated by the cell viability assay in vitro.

\section{Experimental details}
\subsection{Ion implantation and surface characterization}
The AZ31 Mg alloy (Mg with Al 3.12 wt.\%, Zn 0.93 wt.\%, and Mn 0.30 wt.\%) was cut into blocks with dimensions of $10 mm \times 10 mm \times 2 mm$.
The samples were ground with different grit sizes of SiC sandpapers (up to 2000 grit), ultrasonically cleaned in acetone, absolute ethanol for 10 min respectively, and dried in the air.
NH$_{2}^{+}$ ion implantation was performed on an ion implanter equipped with gaseous NH$_{3}$ as the ion source.
The pressure in the target chamber was maintained around $10^{-3}$ Pa, and the current density of ion beam was less than 0.01 A/cm$^{2}$.
The NH$_{2}^{+}$ ions identified by mass spectrometry were accelerated and implanted into the target samples with doses of $1\times10^{16}$, $5\times10^{16}$, and $1\times10^{17}$ ions/cm$^{2}$ at an energy of 50 KeV.

The elemental depth profiles and chemical compositions of the treated samples were determined by X-ray photoelectron spectroscopy with Al K$\alpha$ X-rays as the radiation source (XPS, PHI Quantera II, Ulvac-Phi Inc.).
The approximated sputtering rate for depth profiling analysis was 8.8 nm/min based on the reference standard.
Atomic force microscopy (AFM, SPM-960, Shimadzu, Japan) was used to examine the surface morphology before and after the ion implantation.
The water contact angle was measured by a contact angle goniometer (FTA 200, Dataphysics Inc., USA).
The mechanical properties of the sample surface were determined using Nano Indenter (XP, MTS Systems Corporation, USA).

\subsection{Electrochemical tests}
The electrochemical tests were operated on the electrochemical workstation of a three-electrode cell system with a platinum foil as the counter electrode, a saturated calomel electrode (SCE) as the reference electrode, and the sample as the working electrode.
The exposed surface area of the sample to the Hank$^{\prime}$s solution (NaCl 8.00 g/L, KCl 0.40 g/L, CaCl$_{2}$ 0.14 g/L, MgCl$_{2}$$\cdot$6H$_{2}$O 0.10 g/L, MgSO$_{4}$$\cdot$7H$_{2}$O 0.10 g/L, KH$_{2}$PO$_{4}$ 0.06 g/L, Na$_{2}$HPO$_{4} \cdot $2H$_{2}$O 0.06 g/L, NaHCO$_{3}$ 0.35 g/L, glucose 1.00 g/L) was 1 cm$^{2}$ at the temperature of $37^{\circ}$C.
The potentiodynamic polarization curves were obtained at a scanning rate of 1 mV/s.
The corrosion potential ($E_{corr}$) and the corrosion current density ($I_{corr}$) were measured from the polarization curve according to Tafel analysis.
The electrochemical impedance spectroscopy (EIS)
was performed in a range from 100 kHz to 100 mHz with a 5 mV amplitude perturbation.
For each type of samples, at least three samples were tested.

\subsection{Immersion tests}
Immersion tests on the NH$_{2}$-implanted and the untreated  AZ31 Mg alloys were carried out in Hank$^{\prime}$s solution according to ASTM-G31-72 \cite{ASTM}.
The samples were submerged in the solution with the volume to area ratio of 20 mL/cm$^{2}$ at a temperature of $37^{\circ}$C using water bath.
After immersion for 3, 7, 15, 31 days, the samples were taken out from the solution.
The corrosion products on the samples were removed with chromic acid.
The samples were gently rinsed with distilled water, and then dried in open air.
The changes on the surface morphology were detected using scanning electron microscopy (SEM, S-4800, Hitachi, Japan).
For each immersion time node, the average corrosion rate for 3 parallel specimens was calculated as follows:
\begin {equation}
v=\frac{\Delta m}{St}
\end {equation}
where $v$ is the corrosion rate expressed in units of grams per square meter per hour ($g/(m^{2}\cdot h)$). $\Delta m$ represents the mass loss. $S$ is the surface area of the sample before immersion. $t$ is the immersion time.

\subsection{Cytotoxicity tests}
Mouse MC3T3-E1 preosteoblasts (acquired from American Type Culture Collection (ATCC)) were utilized to evaluate the cytotoxicity of untreated and NH$_{2}^{+}$ implanted AZ31 Mg alloy by indirect cell assay.
Cells were cultured in Dulbecco's Modified Eagle's Medium (DMEM, Gibco) with 10\% fetal bovine serum (FBS) at 37$^{\circ}$C in a humidified
atmosphere of 5\% CO$_{2}$.
Prior to the cell viability experiment, samples were sterilized under ultraviolet radiation for 2 hrs.
Extracts were produced by immersing the samples into serum-free DMEM in a humidified atmosphere of 5\% CO$_{2}$ for 72 hrs.
The surface area of extraction medium ratios were set at 0.5 and 1.25 cm$^{2}$/ml, respectively.
Then the supernatant fluid was withdrawn, filtered, and refrigerated at 4 $^{\circ}C$ for the following assay.
MC3T3-E1 cells were seeded on 96-well culture plates with a density of $1 \times 10^{3}$ cells per well and incubated for 24 hrs to allow attachment.
The medium was then replaced by the corresponding alloy extracts with 10\% FBS.
The DMEM medium with 10\% FBS was set as control groups.
After 1 and 3 days of incubation, 10 $\mu$L MTT was added into each well and then cells were incubated for 4 hrs.
100 $\mu$L dimethyl sulfoxide was added to dissolve the formed formazan crystals.
The absorbance of each well was measured with a microplate reader at the wavelength 490 nm (Cytation3, Bio-Teh, USA).
In the end, the cell viability was computed as follows:
\begin {equation}
Viability = \frac{OD_{sample}}{OD_{control}}\times 100 \%
\end {equation}

\section{Results}

\subsection{Surface characterization}
\begin{figure}
	\centering
		\includegraphics[scale=.80]{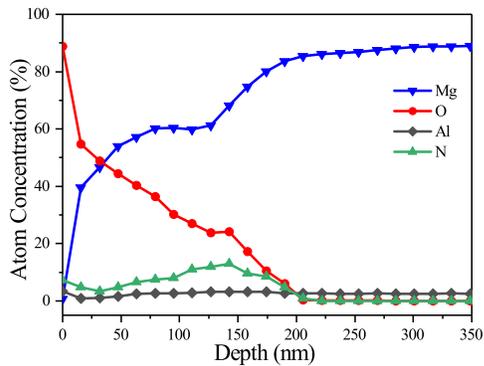}
	\caption{XPS depth profiles of the implanted sample with dose of $1\times10^{17}$ NH$_{2}^{+}$/cm$^{2}$ by XPS.}
	\label{Fig1}
\end{figure}

\begin{figure*}
	\centering
		\includegraphics[scale=1.0]{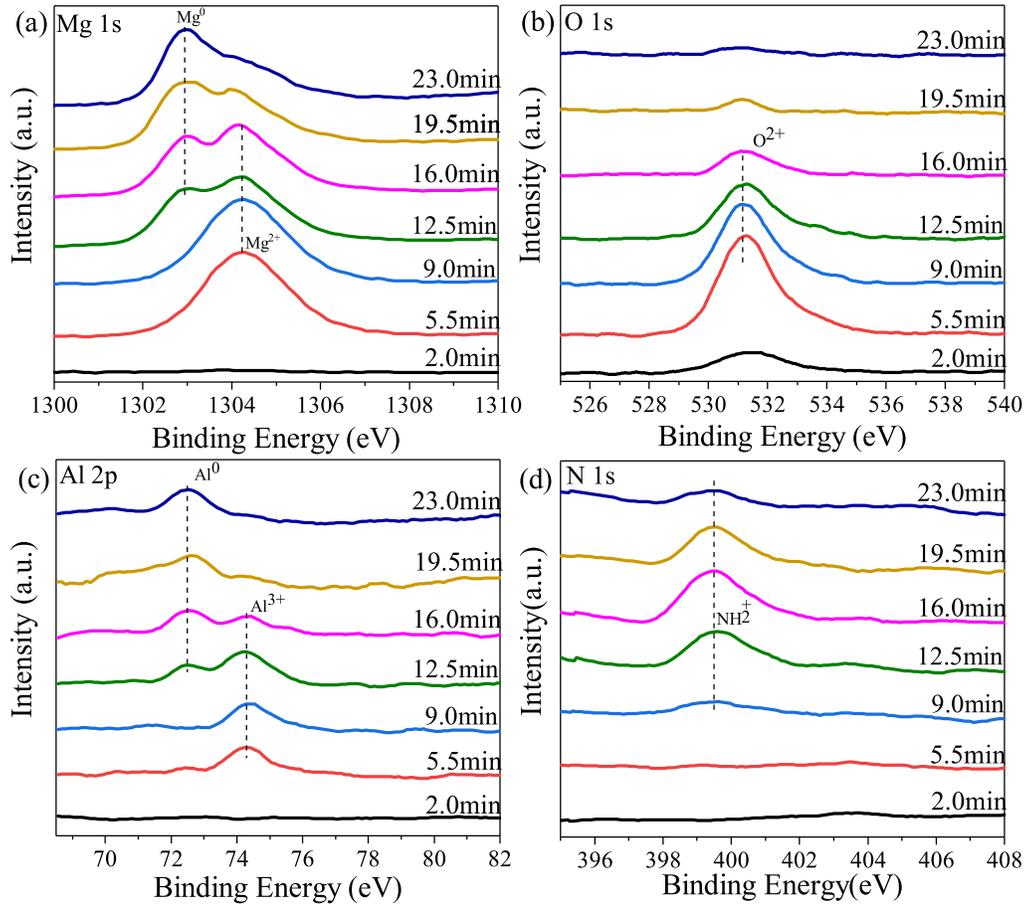}
	\caption{High-resolution XPS spectra of the implanted sample with dose of $1\times10^{17}$ NH$_{2}^{+}$/cm$^{2}$ at different sputtering time.}
	\label{Fig2}
\end{figure*}

The XPS depth profile of the sample implanted with the dose of $1\times10^{17}$ NH$_{2}^{+}$/cm$^{2}$ is depicted in Fig.\ref{Fig1} and the high-resolution XPS Mg 1s, O 1s, Al 2p, and N 1s spectra are shown in Fig. \ref{Fig2}.
As shown in Fig.\ref{Fig1}, the concentration of N atom increases first with the sputtering time and then slowly decreases to 0\%, indicating the formation of a thin NH$_{2}$-containing layer with a maximum concentration of approximately 16\% at a depth of nearly 141 nm.
The existence of oxygen in the top layer results from the formation of surface oxides due to the non-high vacuum condition.
The oxygen and nitrogen concentrations fall down to very low levels at the depth exceeding 202 nm.
The high-resolution Mg 1s spectrum in the Fig \ref{Fig2}(a) shows that Mg is observed in the form of oxidized state (MgO/Mg(OH)$_{2}$) in the top surface layer.
With the longer sputtering time, the Mg 1s peak shifts toward smaller binding energy, implying that Mg gradually turns to the metallic state (Mg$^{0}$).
The binding energy of Al 2p at 74.4 eV suggests that Al$_{2}$O$_{3}$ is observed on the surface.
The intensity of the Al peak shifts from the oxidized state to the metallic state with increasing in depth.
The binding energy of N 1s in the spectrum is around at 398.4 eV indicating the presence of amino.
This result confirms the successful grafting of organic functional groups on the surface of AZ31 magnesium alloys.
As shown in Fig.\ref{Fig2}(b), the oxygen peak represents the formation of MgO, Mg(OH)$_{2}$, and Al$_{2}$O$_{3}$ in the top layer.
The oxygen peak intensity decreases with sputtering time, which is in accordance with the gradually reduced oxygen concentration as shown in Fig. \ref{Fig1}.
The formation of the amino layer and the oxide layer (MgO, Mg(OH)$_{2}$, and Al$_{2}$O$_{3}$) provides a barrier to resist the corrosion of magnesium alloy \cite{JinWXNd, Jameshmiw, WangInvitro, WongInvivo}.

\begin{figure*}
	\centering
		\includegraphics[scale=1.0]{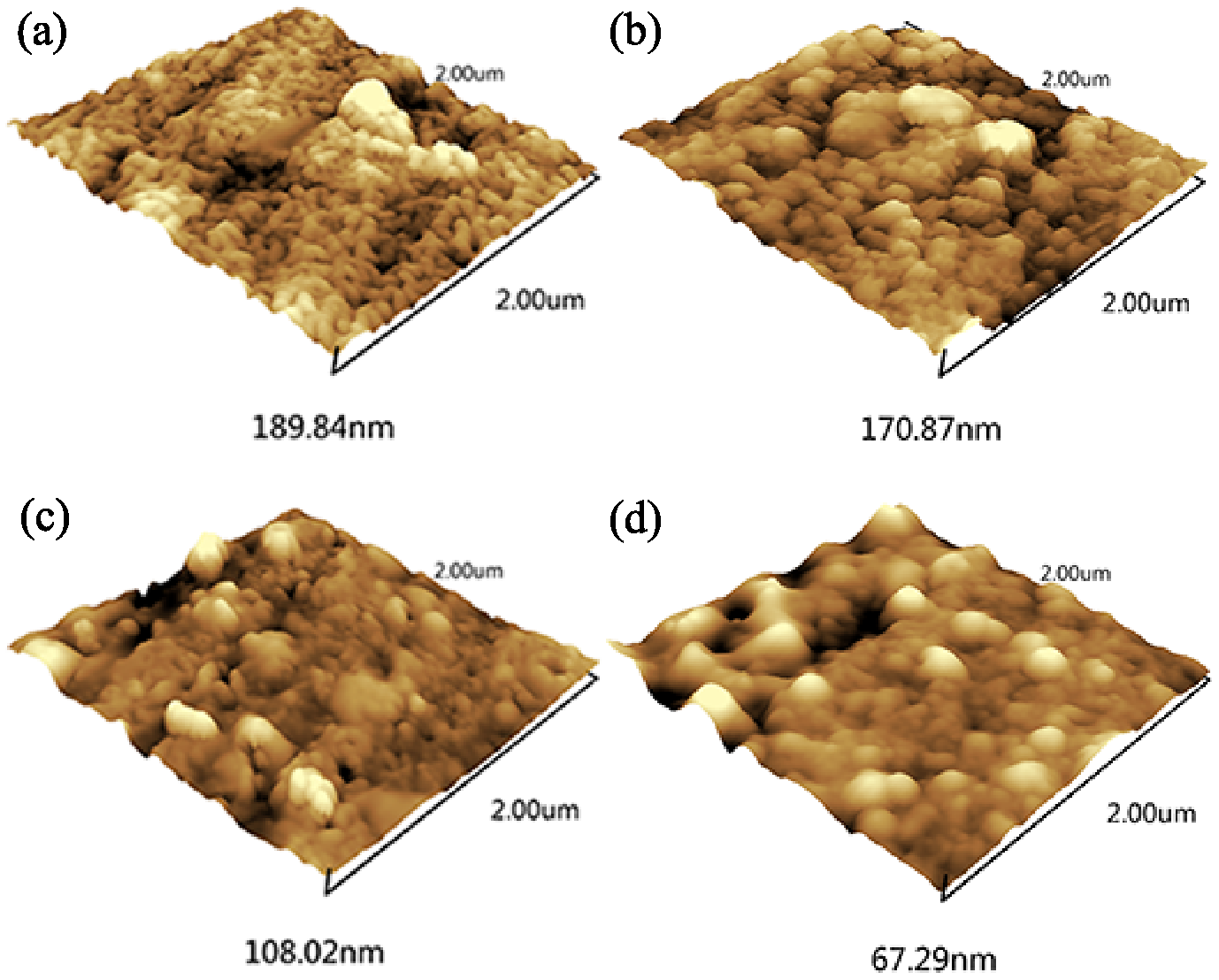}
	\caption{Surface morphology of the samples observed by AFM: (a) untreated, (b) implanted sample with dose of $1\times10^{16}$ ions/cm$^{2}$, (c) implanted sample with dose of $5\times10^{16}$ ions/cm$^{2}$, and (d) implanted sample with dose of $1\times10^{17}$ ions/cm$^{2}$.}
	\label{Fig3}
\end{figure*}

The atomic force microscopy (AFM) images of the untreated and treated samples are shown in Fig. \ref{Fig3}, which reveals the changes in the surface topography after implanted with different doses of NH$_{2}^{+}$ ions.
The surface of the untreated sample is relatively rough with many particles.
After the implantation with various doses of NH$_{2}^{+}$ ions, more compact and finer particles are observed on the surface.
The number of pits is significantly reduced with the higher implantation dose.
The root-mean-square (RMS) roughness values reduce with increasing doses as shown in Table \ref{Table1}.
This indicates that the surface becomes smoother after NH$_{2}^{+}$ ion implantation which may be attributed to the surface restructure at raised temperature during surface modification process \cite{XuEletro}.
Moreover, as reported in literatures \cite{LiWLiDY, HongTNagumeff, SasakiKB}, the smoother surface corresponds to the lower corrosion rate because the electrode potential has small difference between the concave and convex.
Since a relatively smooth surface of sample is obtained by surface modification through NH$_{2}^{+}$ ions, the
sample is expected to be better resistant to corrosion.

\begin{table*}
\caption{Comparison of AZ31 with and without implantation in terms of surface roughness.}\label{Table1}
\setlength{\tabcolsep}{0 mm}{
\begin{tabular}{cccc}
\hline
Doses (ions/cm$^{2}$)& RMS roughness (nm)& $E_{corr}$ (V) & $i_{corr}$ ($\mu$A/cm$^{2}$)\\
\hline
0 & $35.80\pm1.03$ & -1.03 & 869.72\\
$1\times10^{16}$ & $29.68\pm4.31$ & -1.03  &  441.34 \\
$5\times10^{16}$& $17.20\pm1.46$  &  -1.05 & 206.00 \\
$1\times10^{17}$ & $12.80\pm0.44$ &  -1.01 & 125.23\\
\hline
\end{tabular}}
\end{table*}

\begin{figure}
	\centering
		\includegraphics[scale=.8]{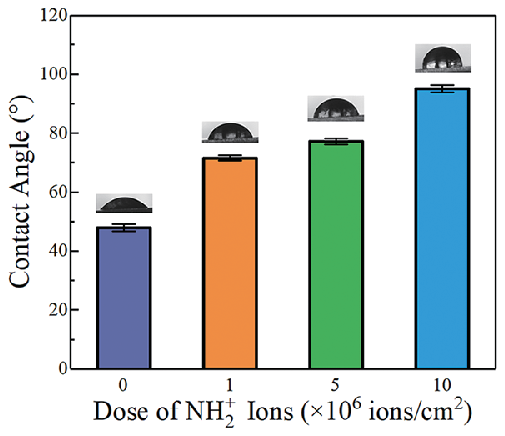}
	\caption{Contact angles for water on AZ31 Mg alloy.}
	\label{Fig4}
\end{figure}

Fig. \ref{Fig4} shows the variation in water contact angles of AZ31 Mg alloys before and after being treated by NH$_{2}^{+}$ ions.
AZ31 Mg alloy is a type of hydrophilic material with a water contact angle of 47.95$^{\circ}$.
With the higher implantation doses, the water contact angle of the treated sample increases.
The higher contact angle corresponds to the lower wettability, which indicates that more hydrophobic surface is obtained after the NH$_{2}^{+}$ ion implantation \cite{SalmanSASelf, WangSuper}.

\subsection{Mechanical property}

\begin{figure*}
	\centering
		\includegraphics[scale=1.0]{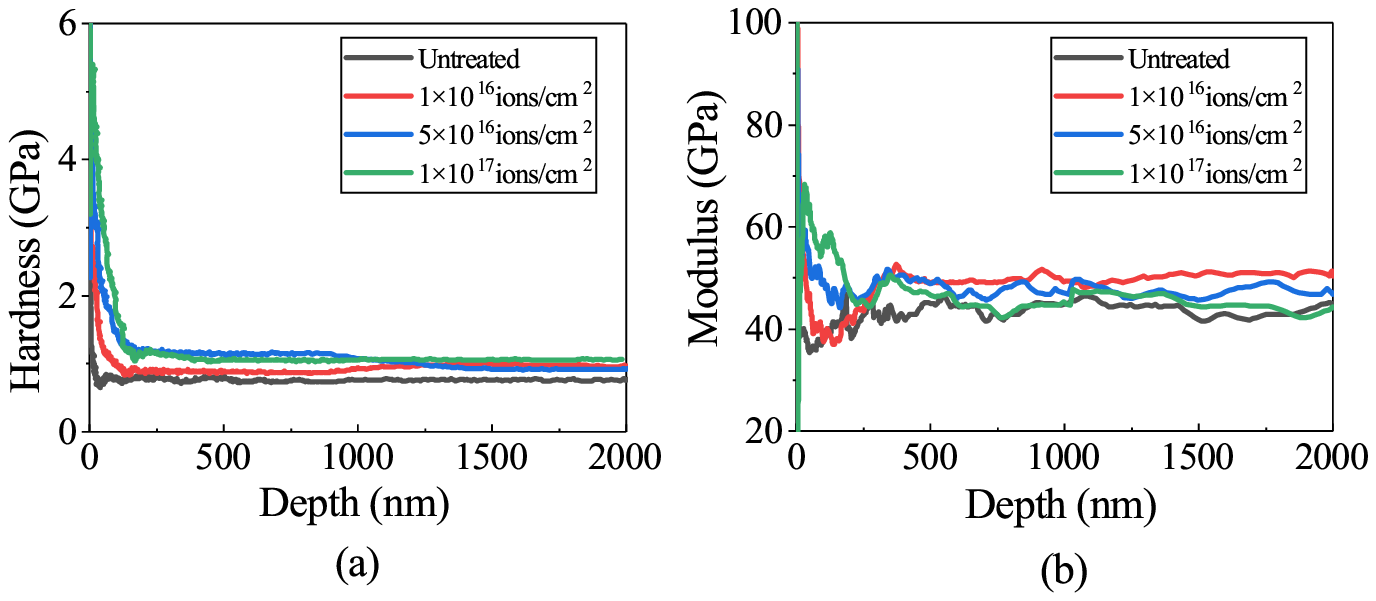}
	\caption{The hardness (a) and the elastic modulus values (b) acquired from the untreated and NH$_{2}^{+}$-implanted samples.}
	\label{Fig10}
\end{figure*}

The results of the hardness and the elastic tests are shown in Fig. \ref{Fig10}.
The indentation depth is much larger than the implantation depth as revealed in the XPS depth profile results, indicating that the data of hardness and the data of elastic modulus are not only based on the implanted layer but also the substrate layer.
Both the hardness and the elastic modulus of the untreated sample maintain a plateau with increasing depth.
While the hardness and elastic modulus of the treated samples exhibit diverse trend with the displacement into the surface.
As to the implanted sample with dose of $1\times10^{17}$ NH$_{2}^{+}$/cm$^{2}$, the maximum hardness reaches 5.4 GPa at 35.7 nm under the surface and then decreases to a plateau value of 1.2 GPa at approximately 1017.7 nm.
The maximum value of elastic modulus is 67.3 GPa at the topmost layer and then decreases to a constant value of 44.5 GPa.
Similar trend can be observed for the samples with doses of $1\times10^{16}$ and $5\times10^{16}$ NH$_{2}^{+}$/cm$^{2}$, respectively.
Both the hardness and the elastic modulus of the treated samples decrease gradually to constant values which are the values of the substrate with increased indentation depth.
Thus, the mechanical performance of the implanted layer is better than the substrate, or at least comparable to the latter \cite{Pooncarbon, ZhaoNano}.

\subsection{Corrosion performance}

\begin{figure}
	\centering
		\includegraphics[scale=.8]{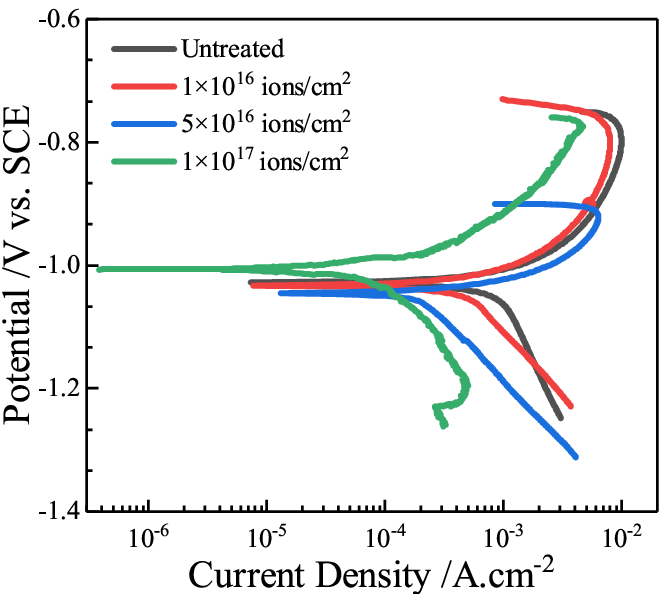}
	\caption{Potentiodynamic polarization curves of untreated and NH$_2^+$ implanted AZ31 in Hank$^{\prime}$s solution.}
	\label{Fig6}
\end{figure}

The potentiodynamic polarization curves of different samples soaked in Hank$^{\prime}$s solution at $37^{\circ}$C are exhibited in Fig. \ref{Fig6}.
The electrochemical parameters are analyzed to characterize the corrosion nature as listed in Table \ref{Table1}.
In contrast to untreated AZ31 magnesium alloy, the samples implanted with NH$_{2}^{+}$ all achieve lower corrosion current density (I$_{corr}$).
The I$_{corr}$ values for the treated samples are shifted towards negative direction as increasing implantation dosage.
This indicates that the surface layer containing more NH$_{2}^{+}$ ions may be more beneficial to restrain the transfer of Cl$^{-}$ during the polarization process \cite{NakatsugawaImp, AkavipatEffect, LeitaoElec}.
Therefore the degradation of magnesium alloy could be suppressed by NH$_{2}^{+}$ ions implantation.

\begin{figure*}
	\centering
		\includegraphics[scale=1.0]{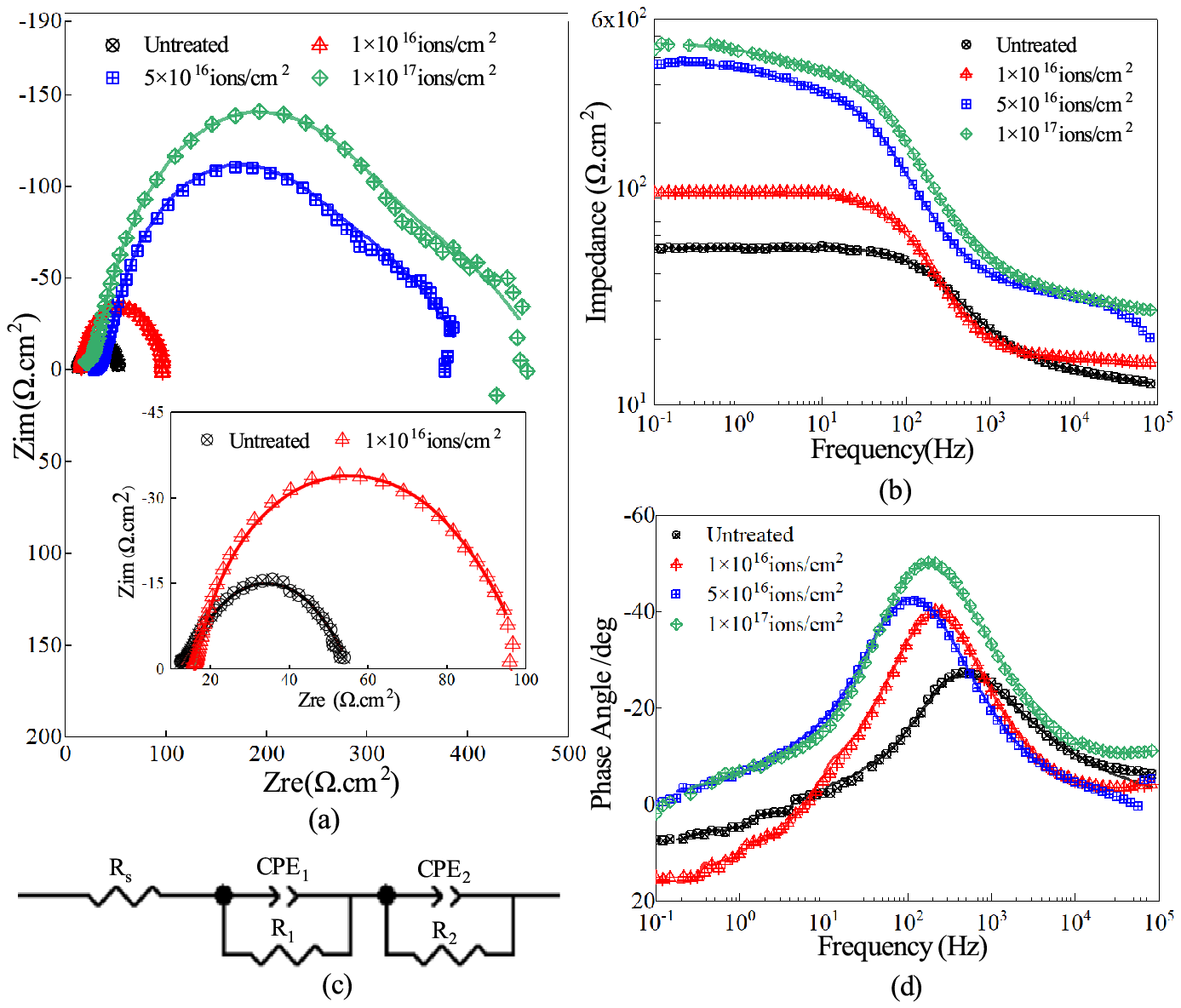}
	\caption{EIS measurements (scatter plot) and model fit (solid lines) of samples: (a) Nyquist plot, (b) Bode plot of impedance versus frequency, (c) Equivalent circuit of samples with and without implantation, and (d) Bode plot of phase angle versus frequency.}
	\label{Fig5}
\end{figure*}

The EIS experiment is conducted to analyze the corrosion property of electrode system.
In general, the impedance spectra are described either by a Nyquist (complex plane) plot in which the opposite of imaginary part of impedance (Zim) versus the real part or by a Bode plot where the modulus of the impedance and the phase angle versus frequency, respectively \cite{GiraultInflu, MacdonaldIm, ChangYY, zhu2018corrosion}.
Fig. \ref{Fig5} shows the EIS data of unimplanted and NH$_{2}$-implanted samples in Hank$^{\prime}$s solution at $37^{\circ}$C.
As shown in Fig. \ref{Fig5}(a), two capacitive loops are emerged at high frequency due to the formation of magnesium hydroxide/oxide and the dissolution of Mg through cracks \cite{CorreaCorr}.
The capacitive loops are distinguishable for the treated samples with dose of $5\times10^{16}$ and $1\times10^{17}$ ions/cm$^{2}$, whereas almost overlapping for the treated sample of $1\times10^{16}$ ions/cm$^{2}$ and untreated sample \cite{NabizadehCom}.
The capacitive loops of the implanted samples are significantly larger than that of the untreated sample.
And the diameter of the capacitive semicircle is the largest for implanted sample with a dose of $1\times10^{17}$ NH$_{2}^{+}$/cm$^{2}$.
Since the larger is the capacitive loop, the better is the corrosion resistance.
The Nyquist plot implies that the anti-corrosion performance of AZ31 Mg alloy improves after NH$_{2}^{+}$ ion implantation.

The Bode impedance plots of the untreated and the implanted sample are displayed in Fig. \ref{Fig5} (b).
The impedances of the implanted samples are higher at all frequency range compared to that of the untreated samples.
As to the sample with high implantation dose ($1\times10^{17}$ NH$^{+}_{2}$/cm$^{2}$), the impedance is 8 times larger than that of the untreated sample at the low frequency of 100 mHz, and the impedance is increased evidently by 2 times compared with that of the untreated sample at the high frequency of 100 kHz.
As reported in literatures \cite{XuEletro, ChenYLong, AmirudinApp, MurrayJN, ShiA, ParkJHL}, the higher impedances at low and high frequencies reflect the slower process of charge transfer and the poor ability of the electrolyte to penetrate.
So the Bode impedance result implies that surface protection is enhanced after NH$_{2}^{+}$ ion implantation.
The Bode phase angle evolution of the untreated and the implanted sample is shown in Fig. \ref{Fig5} (d).
The maximum phase angles of treated samples increase with increasing implantation dose, and are all larger than that of the untreated sample.
More capacitive behavior is demonstrated by larger phase angle, which suggests the occurrence of a more stable and denser layer to retard the electrolyte penetration to the substrate \cite{WangFlu, Deassiss}.
Hence, the difficulties in charge transfer and electrolyte penetration confirm that ion implantation effectively improves the corrosion resistance.

To further investigate the enhanced corrosion resistance, the electrical equivalent circuit is utilized to fit the EIS spectra of the untreated sample and the implanted sample, as shown in Fig. \ref{Fig5} (c).
The equivalent circuit displays two layers, a porous outer layer and a barrier-like inner layer.
CPE$_{1}$ is the constant phase element of the outer porous layer and R$_{1}$ is the corresponding resistance.
The CPE$_{2}$ stands for the constant phase element in the inner layer in parallel connection with the resistance R$_{2}$.
R$_{s}$ represents the solution resistance between the working electrode and the reference electrode.
The fitted parameters of the electrical equivalent circuit are listed in Table \ref{Table2}.
As disclosed in Table \ref{Table2}, R$_{1}$ and R$_{2}$ increase while CPE$_{1}$ and CPE$_{2}$ decrease after samples implanted by NH$_{2}^{+}$ ion, indicating that the corrosion rates are retarded in Hank$^{\prime}$s solution.
As to the sample of high implantation dose ($1\times10^{17}$ ions/cm$^{2}$), it achieves the highest resistance, approximately 10 times in R$_{1}$ and R$_{2}$ larger than that of the untreated sample.
The resistance R$_{2}$ of the inner layer is dramatically increased which is attributed to the formation of MgO, Al$_{2}$O$_{3}$, and amino based on the results of the XPS depth profile.
Hence, the stable inner layer of the implanted sample can further reinforce the corrosion resistance with the corrosion evolution.
These EIS results suggest that the samples modified by NH$_{2}^{+}$ ion improve the anti-corrosion property of AZ31 Mg alloys and are in accordance with the polarization curve.

\begin{table*}
\caption{\label{Table2}Electrochemical parameters of AZ31 with and without implantation obtained by equivalent circuit
simulation.}
\setlength{\tabcolsep}{0 mm}{
\begin{tabular}{cccccccc}
\hline
Doses (ions/cm$^{2}$)& R$_{s}$($\Omega cm^{2}$) &CPE$_{1}$($\Omega ^{-2}cm^{2}S^{-n}$)& $n_{1}$ &$R_{1}(\Omega cm^{2}$)&CPE$_{2}(\Omega^{-2}cm^{-2}S^{-n}$)& $n_{2}$& $R_{2}(\Omega cm^{2})$\\
\hline
0 & 11.83 & $3.93\times10^{-5}$ & 0.98 & 19.74 & $7.42\times10^{-4}$ & 0.54 & 24.59 \\
 $1\times10^{16}$ & 15.67 & $3.64\times10^{-5}$ & 0.98 & 50.78 & $6.64\times10^{-4}$ & 0.67 & 34.11 \\
 $5\times10^{16}$ & 35.52 & $2.77\times10^{-5}$ & 0.96 & 159.00 & $5.55\times10^{-4}$ & 0.59 & 207.40\\
 $1\times10^{17}$ & 21.93 & $1.55\times10^{-5}$ & 0.99 & 197.10 & $5.09\times10^{-4}$ & 0.54 & 262.20\\
\hline
 \end{tabular}}
\end{table*}

\begin{figure}
	\centering
		\includegraphics[scale=.8]{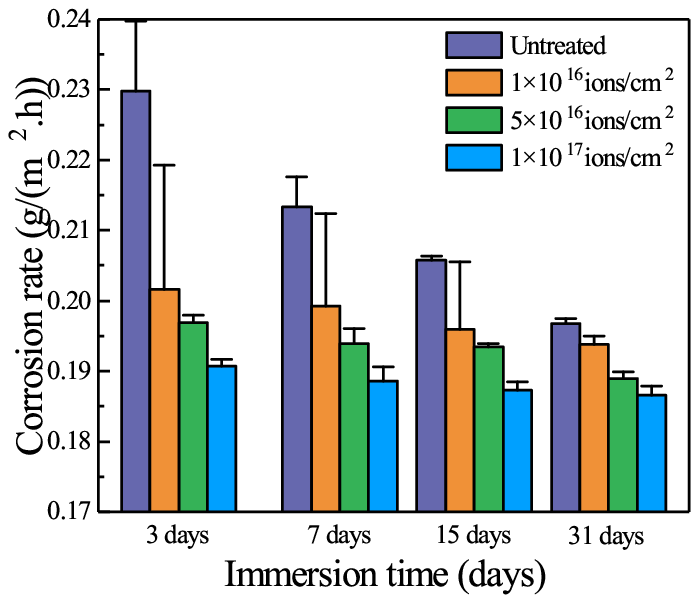}
	\caption{The corrosion rate of samples with and without implantation during different immersion time.}
	\label{Fig7}
\end{figure}

\begin{figure*}
	\centering
		\includegraphics[scale=1.0]{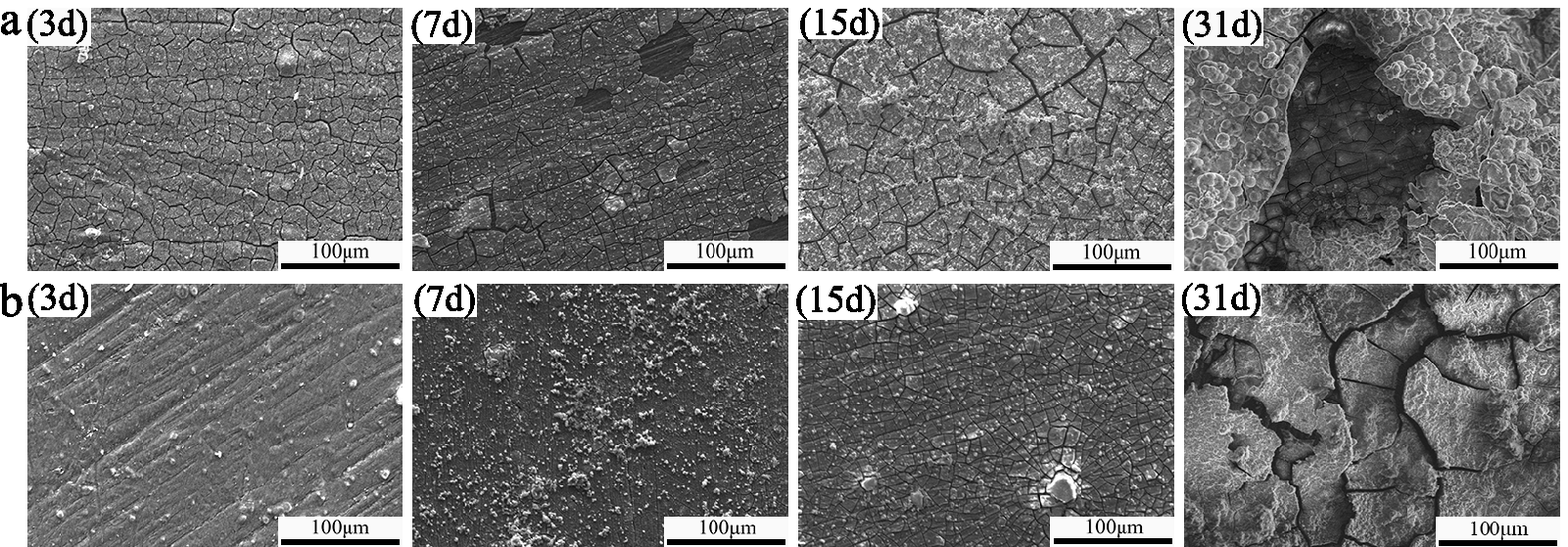}
	\caption{Surface morphology of samples in different immersion time periods. Group (a) is the bare sample, and group (b) is implanted sample with dose of $1\times10^{16}$ ions/cm$^{2}$.}
	\label{Fig8}
\end{figure*}

The immersion test is conducted to further study the corrosion resistance behavior.
The corrosion rate is calculated by the mass loss method with sample soaked in the Hank$^{\prime}$s solution for 3, 7, 15, 31 days at 37$^{\circ}$, respectively.
As shown in Fig. \ref{Fig7}, the corrosion rate decreases with increasing immersion time.
The corrosion rates of NH$_{2}$-implanted samples have smaller values than that of the untreated sample.
Fig. \ref{Fig8} exhibits the corrosion morphologies of the NH$_{2}$-implanted and the untreated samples after immersed into the Hank$^{\prime}$s solution for 3, 7, 15, 31 days at $37^{\circ}$C, respectively.
Group (a) is the surface morphology of the untreated sample while Group (b) is the surface morphology of the NH$_{2}$-implanted sample with dose of $1\times10^{16}$ ions/cm$^{2}$.
As shown in Group (a) of Fig. \ref{Fig8}, cracks are obviously observed after immersion for 3 days.
The general trend of crack spacing goes bigger along with longer immersion time.
The untreated sample even suffers from severe corrosion pits after immersion for 31 days.
On the contrary, some tiny cracks are observed on the NH$_{2}$-implanted sample after immersion for 3 and 7 days.
Although the corrosion cracks grow larger, the obvious corrosion pit does not appear on the treated surface after immersion for 31 days in Group (b).
Therefore, the corrosion rate of the AZ31 Mg alloy is effectively decreased after the NH$_{2}^{+}$ ion implantation.

\subsection{In vitro cytotoxicity studies}
\begin{figure}
	\centering
		\includegraphics[scale=.75]{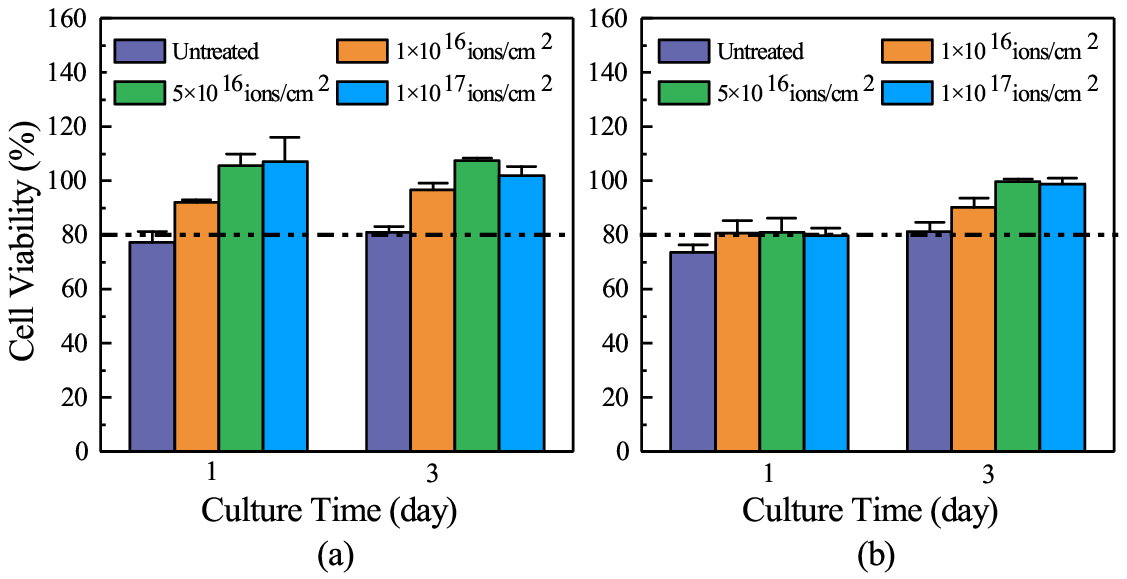}
	\caption{In vitro viability of MC3T3-E1 cells cultured in unimplanted and implanted AZ31 extraction
mediums for 1 and 3 days with different extraction medium ratios: (a) 0.50 cm$^{2}$/ml and (b) 1.25 cm$^{2}$/ml.}
	\label{Fig9}
\end{figure}

In order to evaluate the vitro cytotoxicity of the samples, the viability of MC3T3-E1 cells with different extraction medium ratios (0.50 and 1.25 cm$^{2}$/ml) after incubation for 3 day is measured, as shown in Fig. \ref{Fig9}.
The cell viability of the control group is considered as 100\%.
As known in ISO 10993-5, cell viability is larger than 80\% corresponding to Grade 0 or Grade 1 of cytotoxicity of materials, which indicates that this material can be used as biomaterials \cite{BianDevep, ISO109}.
For the extraction medium ratio of 0.5 cm$^{2}$/ml, cells of the untreated samples exhibit viabilities below 80\%, indicating obvious cytotoxic effects according to the ISO 10993-5.
However, cell viabilities in NH$_{2}^{+}$-AZ31 extracts are all above 80\% and higher than that in the untreated AZ31 extract during the whole incubation period.
As to the extraction medium ratio of 1.25 cm$^{2}$/ml (Fig. \ref{Fig9}(b)), the cells display roughly the same viability (about 80\%) for different samples after incubation for 1 day.
However, after 3 days of incubation, cells grown in the extract from NH$_{2}^{+}$-implanted samples present significantly increased activity compared to the untreated sample, indicating that the cytotoxic effect of the implanted sample on MC3T3-E1 preosteoblasts is not developed, though the reduced cell viability is observed in the first day of incubation.
This indicates that the implanted samples with the smoother surface and the reduced corrosion rate show the relatively stable and biofriendly effect on cell growth \cite{Sundep}.
Overall, the cell viability of samples with NH$_{2}^{+}$ ion implantation is significantly higher than that of the untreated samples after the incubation for 3 days, indicating the better biocompatibility of the NH$_{2}^{+}$-implanted AZ31 Mg alloy \cite{BagherifardSSP}.

\section{Discussion}

\subsection{Corrosion behavior}
According to the XPS spectra (Fig. \ref{Fig1}-\ref{Fig2}), NH$_{2}^{+}$ ions have been successfully implanted into the sample surface with the formation of the ammonia and the oxide layers.
A relatively smooth and compact surface of treated sample is observed by AFM (Fig. \ref{Fig3}).
As shown in Fig. \ref{Fig4}, a larger water contact angle of the treated sample suggests that a hydrophobic surface is obtained after the NH$_{2}^{+}$ ions implantation.

The modulated smooth surface with hydrophobicity significantly improves the corrosion resistance of AZ31 Mg alloy which can be proved by the electrochemical tests (Fig. \ref{Fig6}-\ref{Fig5}) and the immersion test (Fig. \ref{Fig7}-\ref{Fig8}) in Hank$^{\prime}$s solution at the temperature of $37^{\circ}$C.
The degradation process is generally accompanied by electrochemical reactions.
Magnesium is one of the active metals that can quickly dissolve (Eq. (\ref{eq:11})) at the anode electrode during immersion \cite{SongUn, SongInfl}.
Mg$^{2+}$ reacts with H$_{2}$O to generate hydrogen gas at the cathode electrode (Eq. (\ref{eq:22})).
The Mg(OH)$_{2}$ layer forms with consuming the H$^{+}$ and producing the OH$^{-}$ (Eq. (\ref{eq:33})), creating an alkaline environment.
\begin{equation}
\begin{aligned}
Mg \rightarrow Mg^{2+}+2e^{-}
\label{eq:11}
\end{aligned}
\end{equation}
\begin{equation}
\begin{aligned}
H_{2}O + 2e^{-} \rightarrow H_{2}+2OH^{-}
\label{eq:22}
\end{aligned}
\end{equation}
\begin{equation}
\begin{aligned}
Mg^{2+} + 2OH^{-} \rightarrow Mg(OH)_{2}
\label{eq:33}
\end{aligned}
\end{equation}
The Mg(OH)$_{2}$ can be corroded by Hank$'$s solution which contains abundant Cl$^{-}$.
According to the penetration/dissolution mechanism as reported in \cite{DuanElectro}, Cl$^{-}$ disperses and interacts in the hydroxide, then penetrates through the hydroxides to reach the substrate, which results in the dissolution of the substrate.
The thinner Mg(OH)$_{2}$ layer of the untreated sample is easily attacked by Cl$^{-}$ leading to the poor corrosion resistance of the material.
However, for samples implanted by NH$_{2}^{+}$, the stable products including the ammonia layer and oxide layer (MgO/Al$_{2}$O$_{3}$) are formed as the strong barriers against the corrosion to the substrate.
These layers can delay the electrolytes to Mg substrate.
Therefore, the corrosion rate is decreased after the NH$_{2}^{+}$ ion implantation.

\subsection{Cytotoxicity evaluation}

Although the AZ31 magnesium alloy is a candidate of biomaterial because of its mechanical property and biodegradability, improvements in chemical stability in living tissues, biological properties including wettability, adhesion and biocompatibility are required so that it could have strong bondage with organic bone tissue.
The -NH$_{2}$ and -NH amidogen radicals are helpful in simulating the activity of living tissues.
In order to evaluate the biocompatibility of NH$_{2}^{+}$ implanted AZ31, we have conducted the toxicity assay which is a vital index for rapidly screening the biocompatibility of biometal materials.
The toxicity of implants is mainly affected by the amount of released ions during degradation.
In practice, cytotoxic effect occurs if the released ion concentration is beyond the tolerance limit.
Moreover the corrosion process contributes to the local alkalization and hydrogen evolution which are harmful to cell viability.

As shown in Fig. \ref{Fig9}, samples implanted by NH$_{2}^{+}$ ions do not induce toxicity to MC3T3-E1 cells after cultured for 3 days.
This indicates that the decreased corrosion rate leads to the limited alkalization and hydrogen evolution and delays the release of ions.
Additionally, the -NH$_{2}$ amidogen radical as a functional group, which is generally found in organic molecules, is less toxic in comparison with the metal ions.
In contrary, the untreated sample presents a lower cell viability compared to the implanted ones, mainly resulting from the higher concentration of the released metal ions.
Therefore, the biocompatibility is enhanced through grafting the -NH$_{2}$ amidogen radical onto the inorganic biomaterial (AZ31).

\section{Conclusions}
In this study, AZ31 magnesium alloy is implanted by different doses of NH$_{2}^{+}$ ions at energy of 50 KeV.
The degradation behavior and the cytotoxicity of the ion implanted samples have been investigated.
The results from potentiodynamic polarization, electrochemical impedance spectroscopy, and immersion tests demonstrate that the implanted samples show the improved performance in corrosion resistance.
This is attributed to the formation of a stable and protective layer composed of oxide and amino with stable chemical bonds as a barrier to prevent the material from being corroded by biological liquid.
Furthermore, the modification by NH$_{2}^{+}$ ions improves the bioactivity of the material surface.
In the toxicity experiment, higher viability of implanted samples is observed from the extraction medium ratio of 1.25 cm$^{2}$/ml after incubating for 3 days, which confirms the desired biocompatibility of the treated samples in vitro.
Our data indicates that NH$_{2}^{+}$ ion implantation is a favorable method of improving corrosion resistance and cytocompatibility of the AZ31 magnesium alloy.
The method of implanting simple organic functional groups on Mg alloy surface may provide a new perspective for the surface treatment for biomaterials.

\section*{Acknowledgments}
This work is supported by National Science Foundation (NSF) of China with the Grant No.$11675014$.
Additional support is provided by Ministry of Science and Technology of China $(2013YQ030595-3)$.





\bibliographystyle{model1-num-names}

\bibliography{ref1}
\end{document}